\documentclass[fleqn,usenatbib]{mnras}

\usepackage{newtxtext,newtxmath}

\usepackage[T1]{fontenc} 

\DeclareRobustCommand{\VAN}[3]{#2}
\let\VANthebibliography\thebibliography
\def\thebibliography{\DeclareRobustCommand{\VAN}[3]{##3}\VANthebibliography}

\usepackage{graphicx}	
\usepackage{amsmath}	
\usepackage{amssymb}	
\usepackage{verbatim}

\usepackage{color} 

\title[Bias of the core shift in AGN jets]{A bias in VLBI measurements of the core shift effect in AGN jets}

\author[I. N. Pashchenko et al.]{
I.~N. Pashchenko,$^{1}$\thanks{E-mail:in4pashchenko@gmail.com}
A.~V. Plavin,$^{1,2}$
A.~M.~Kutkin,$^{1,3}$
Y.~Y.~Kovalev$^{1, 2, 4}$
\\
$^{1}$Astro Space Center, Lebedev Physical Institute, Profsouznaya 84/32, Moscow 117997, Russia\\
$^{2}$Moscow Institute of Physics and Technology, Institutsky per. 9, Moscow region, Dolgoprudny, 141700, Russia\\
$^{3}$ASTRON, The Netherlands Institute for Radio Astronomy, Oude Hoogeveensedijk 4, 7991 PD, Dwingeloo, The Netherlands\\
$^{4}$Max-Planck-Institut f{\"u}r Radioastronomie, Auf dem H{\"u}gel 69, D-53121 Bonn, Germany
}

\date{Accepted 2020 October 7. Received 2020 October 7; in original form 2020
July 27}

\pubyear{2020}

\begin{document}
\label{firstpage}
\pagerange{\pageref{firstpage}--\pageref{lastpage}}
\maketitle

\begin{abstract}
The Blandford and K\"{o}nigl model of AGN jets predicts that the position of the apparent opaque jet base --- the core --- changes with frequency. This effect is observed with radio interferometry and is widely used to infer parameters and structure of the innermost jet regions.
The position of the radio core is typically estimated by fitting a Gaussian template to the interferometric visibilities. This results in a model approximation error, i.e. a bias that can be detected and evaluated through simulations of observations with a realistic jet model.
To assess the bias, we construct an artificial sample of sources based on the AGN jet model evaluated on a grid of the parameters derived from a real VLBI flux-density-limited sample and create simulated VLBI data sets at 2.3, 8.1 and 15.4 GHz.
We found that the core position shifts from the true jet apex are generally overestimated. The bias is typically comparable to the core shift random error and can reach a factor of two for
jets with large apparent opening angles.
This observational bias depends mostly on the ratio between the true core shift and the image resolution. This implies that the magnetic field, the core radial distance and the jet speed inferred from the core shift measurements are overestimated.
We present a method to account for the bias.
\end{abstract}

\begin{keywords}
techniques: interferometric -- galaxies: active -- galaxies: nuclei -- galaxies: jets
\end{keywords}


\section{Introduction}

The Very Long Baseline Interferometry (VLBI) technique allows exploring Active Galactic Nuclei (AGN) jets with the highest angular resolution. Synthesised images of most of the radio-loud AGN observed with VLBI are dominated by an unresolved or barely-resolved feature called the core \citep[or VLBI core, e.g.][]{1997ARA&A..35..607Z,2005AJ....130.2473K}. The core has a flat or inverted radio spectrum, characteristic of optically-thick synchrotron emission \citep{2014AJ....147..143H}. Its frequency-dependent position was discovered by \cite{1984ApJ...276...56M} and can be explained within the relativistic jet model \citep{1979ApJ...232...34B,1981ApJ...243..700K}. The core appears in that model as a surface in a continuous flow where the optical depth of jets synchrotron radiation at a given frequency $\tau_\nu \approx 1$~--- ``photosphere''. However, there are other explanations of the core nature \citep[e.g.][]{2008ASPC..386..437M}.

The measurements of the core shift effect are of a key importance in studying AGN jets. They were used to estimate the jet magnetic field and particle density \citep{1998A&A...330...79L,2005ApJ...619...73H,2008A&A...483..759K,2009MNRAS.400...26O,2011A&A...532A..38S,2012MNRAS.420..542A,2012A&A...545A.113P,2013A&A...557A.105F,2014MNRAS.437.3396K,2015MNRAS.451..927Z,2017MNRAS.468.2372N,Finke_2019,2019MNRAS.485.1822P}, the composition of the jet plasma \citep{2005ApJ...619...73H}, the jet magnetization \citep{2015MNRAS.447.2726N}, the jet power and its production efficiency \citep[e.g][]{2017MNRAS.465.3506P,2020IAUS..342..197N}, the location of core relative to the central black hole and the geometry of the outflow \citep[e.g.][]{2011Natur.477..185H,2017ApJ...834...65A}. They were also used to constrain the spins of the supermassive black holes \citep{2017MNRAS.469..813A} and the nature of the central engine \citep{2014Natur.510..126Z}.
If a source demonstrates flaring activity, it is possible to derive the jets innermost regions kinematics based on the measured core shift and multi-frequency time delays \citep{2011MNRAS.415.1631K,2014MNRAS.437.3396K,2018MNRAS.475.4994K,2019MNRAS.486..430K,2019MNRAS.485.1822P}.
Differences in core positions should be taken into account for image registration when constructing the VLBI spectral index \citep[e.g.][]{2001ApJ...550..160M,2008A&A...483..759K,2014AJ....147..143H} and Faraday rotation maps \citep[e.g.][]{2012AJ....144..105H}.
The core shift effect is also important for the radio (VLBI) and optical (Gaia) reference frame alignment \citep{2009A&A...505L...1P,2017A&A...598L...1K,2019ApJ...871..143P}, as it can partially explain the observed offsets.

However, conventionally the position of the core and, thus, the core shift is estimated by fitting a set of Gaussian templates to the interferometric visibility. This could result in a biased estimates of the core shift if the emission profile is non Gaussian. \cite{2019MNRAS.485.1822P} assumed a linear dependence between the measured core shift and the synthesised beam for a multi-epoch multi-frequency (2.3 and 8.4~GHz) sample of 40~sources and found that the measured core shift is positively correlated with resolution, i.e. the beam size. Their results could imply that core shifts are typically overestimated. However, \cite{2011A&A...532A..38S} did not find larger core shifts between 1.4 and 15.4~GHz in the direction of the elongated beam in their sample of 20~sources. In this paper we study the possible bias directly, using the jet model \citep{1979ApJ...232...34B,1981ApJ...243..700K} that is used to explain the core shift effect.

Throughout the paper, we adopt the $\Lambda$CDM cosmology with $\Omega_m=0.287$, $\Omega_\Lambda=0.7185$ and $H_0=69.3$~km~s$^{-1}$~Mpc$^{-1}$ \citep{2013ApJS..208...19H}. The spectral index is defined via the dependence of the flux density on the spectral index $S \propto \nu^{-\alpha}$.

\section{Jet model and core shift measurements}
\label{sec:bkmodel}

The jet model by \cite{1979ApJ...232...34B,1981ApJ...243..700K}
assumes a conical geometry of the outflow and the particle density amplitude and the magnetic field dependence on the distance $r$ from the jet apex: $K = K_1(r/r_1)^{-n}$ and
$B = B_1(r/r_1)^m$, where the reference position is $r_1=1$~pc and $B_1$ and $K_1$ are the values at the $r_1$. In the general case of arbitrary exponents $m$, $n$ and optically thin spectral index $\alpha$ analytical expressions for the distribution of optical depth and brightness temperature can be found in, e.g. \cite{2019MNRAS.488..939P}.


\cite{1998A&A...330...79L} has shown that the relativistic jet model can be used to estimate the jets physical parameters, e.g. magnetic field and particle density, using the measured core shift.
However, this requires some further assumptions because core shift measurements alone do not constrain the parameters of the jet model. More on degeneracies between model parameters can be found in \cite{2019MNRAS.488..939P}. The most widely used is the assumption of an equipartition between emitting particles and magnetic field energy densities \citep{1998A&A...330...79L,2009MNRAS.400...26O,2012A&A...545A.113P}. \cite{2015MNRAS.451..927Z} used the optically thick approximation to avoid the equipartition assumption \citep[for a similar approach, see also][]{2017MNRAS.468.2372N}. They obtained magnetic field estimates which significantly differ from the corresponding equipartition values. Both approaches needed measurements of the core shifts between two frequencies $\triangle r$. In the equipartition case, the magnetic field scales with the core shift as $B_1^{\rm eq} \propto \triangle r^{0.75k_{\rm r}}$ \citep{1998A&A...330...79L}, where $k_r = \frac{n+m(1.5+\alpha)-1}{1.5+\alpha}$. However, for a more general case, the dependence on the measured core shift is much stronger: $B_1^{\rm noeq} \propto \triangle r^{5}$ \citep{2015MNRAS.451..927Z}.


There are several methods to measure the core shift effect \citep[for a brief review, see][]{2014MNRAS.437.3396K}.
In general, one has to know the relative position of the core at several frequencies. The absolute position information is lost during self-calibration of the VLBI data \citep{1984ARA&A..22...97P}. Thus in the absence of a phase calibrator, the first step in deriving the core shift is matching images at different frequencies using a reference point that is stable with frequency. The optically thin structure of a jet is the most obvious reference for core shift measurements. Depending on the morphology of the extended jet structure, two methods are conventionally used. In the first one, the core position is measured with respect to an individual jet feature using a brightness distribution model \citep{1998A&A...330...79L,2008A&A...483..759K,2011A&A...532A..38S,2014MNRAS.437.3396K}. The second measure core position with respect to a large jet section rich
in structure using image cross-correlation
\citep{2000ApJ...530..233W,2008MNRAS.386..619C,2009MNRAS.400...26O,2012MNRAS.420..542A,2012AJ....144..105H,2012A&A...545A.113P,2013A&A...557A.105F,2019MNRAS.485.1822P}.

Regardless of the method used to align images, the final step in deriving the core shift is the measurement of the core position itself. Conventionally, the source structure is modelled with a set of Gaussian templates \citep[e.g.][]{mojave6kinematicsanalysis}. More recently, \cite{2019MNRAS.485.1822P} employed subtraction of the extended structure from the original data set and considered the resulting data as a contribution from the VLBI core only. In both approaches, the core brightness distribution is fitted with a simple model, i.e. usually a circular Gaussian. This could result in a bias of the measured core position if the true brightness distribution is significantly non-Gaussian. Here we define the bias as the difference between the observed and true values. However, before exploring the bias of the measurement, it is worth finding out what is supposed to be measured and what is actually measured.

\section{Bias due to $\tau = 1$ region being offset from the intensity peak}
\label{sec:tauoneorrmax}

To find the analytical expression for the position of the maximum intensity at some frequency, one has to differentiate the equation of the brightness temperature radial distribution.
The brightness temperature distribution in a general case of arbitrary $n$, $m$ and $\alpha$ is \citep{2019MNRAS.488..939P}:
\begin{equation}
\label{eq:Tb}
T(r_{\rm obs}, d) = \frac{c^2 C_{1}(\alpha)D^{0.5}\nu^{0.5}_{\rm obs}r_{\rm obs}^{0.5m}}{8\pi k_B C_2(\alpha) (1+z)^{0.5} B_1^{0.5}r_1^{0.5m}(\sin{\theta})^{0.5m}}(1 - e^{-\tau(r_{\rm obs}, d)})\,,
\end{equation}
where the optical depth is
\begin{multline}
\label{eq:tau}
\tau(r_{\rm obs}, d) =  C_{\rm 2}(\alpha)(1+z)^{-(\alpha+2.5)}D^{1.5+\alpha}K_1 B_1^{1.5+\alpha}\nu_{\mathrm{obs}}^{-(2.5+\alpha)} \\
\times 2r^{n+m(1.5+\alpha)}_{1}\left(\sin{\theta}\right)^{n+m(1.5+\alpha)-1} \sqrt{\phi_{\rm app}^2-\left(\frac{d}{r_{\rm obs}}\right)^2}r_{\rm obs}^{-(n+m(1.5+\alpha)-1)}\,.
\end{multline}
Here the position in the sky is described by $r_{\rm obs}$ and $d$~--- distances along and perpendicular to the jet direction, $D$~--- Doppler factor, $\phi_\mathrm{app}$ is the observed jet half-opening angle, $z$~--- redshift, $C_{1}(\alpha)$ and $C_{2}(\alpha)$ --- functions of optically thin spectral index $\alpha$, $c$ --- speed of light and $k_{\rm B}$ --- Boltzmann constant.

To this end, the following relation for the optical depth at the maximum intensity is obtained:
\begin{equation}
    \tau_{r_{\rm max}} = \log\left(1+\frac{n+m(1.5+\alpha)-1}{0.5m}\tau_{r_{\rm max}}\right).
\label{eq:findtaumax}
\end{equation}
For $n=2$, $m=1$, $\alpha=0.5$ we obtain $\tau_{r_{\rm max}} \approx 2.92$. This stays close to $\tau \approx 3$ for other physically meaningful power-law exponents $n, m$ and spectral index $\alpha$. Thus to obtain the position of the maximum $r_{\rm max}$ for some frequency, one has to solve the equation of the optical depth radial distribution for $\tau = 2.92$.

\begin{figure}
\centering
\includegraphics[width=0.96\columnwidth]{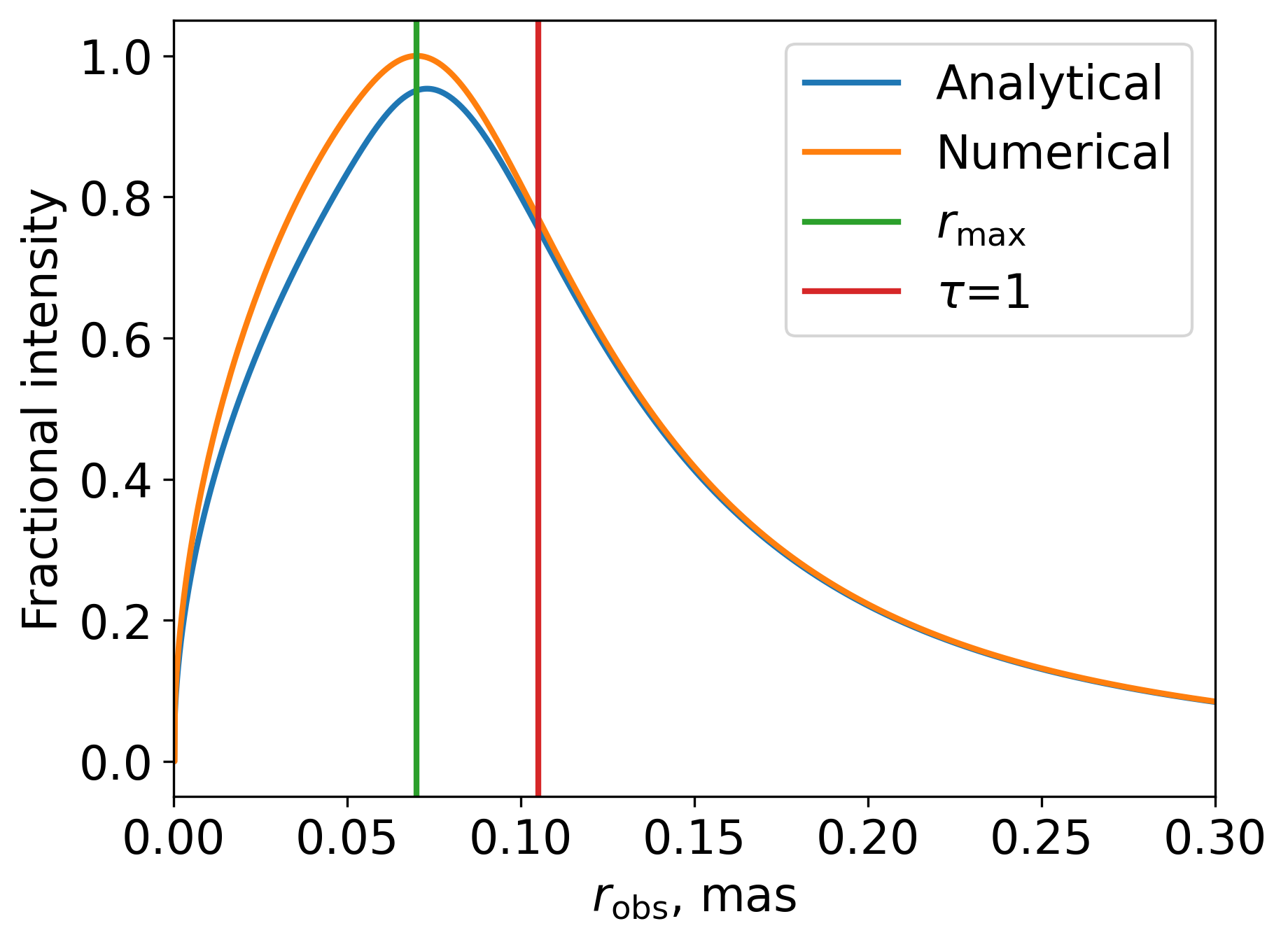}
    \caption{Radial slice of the emission pattern from a Blandford-K\"{o}nigl jet model for viewing angle $\theta = 2^\circ$ and half-opening jet angle $\phi = 0.5^\circ$ showing the difference between the positions of the VLBI core according to its different definition as the region of $\tau = 1$ (red vertical line) and the region of maximal intensity (green vertical line). Both curves are normalised on the maximum of the numerically evaluated radial profile.}
    \label{fig:bkslice}
\end{figure}

Some works \citep[e.g.][]{1979ApJ...232...34B} associate the core with this maximum intensity region, and derive parameter estimates based on that. However, \citep[e.g.][]{1998A&A...330...79L} the VLBI core is often interpreted as the region of unity optical depth, $\tau = 1$. These two approaches are not equivalent, and combining equations from papers adopting different conventions can become confusing and lead to subtle biases.

Both definitions of the core consider the on-axis distributions. An example of the on-axis intensity profile of the Blandford-K\"{o}nigl jet is given in Fig.~\ref{fig:bkslice}. The ratio of the distances from the jet apex to the $\tau=1$ region $r_{\tau{\rm=1}}$ and to the intensity maximum $r_{\rm max}$ can be calculated by substituting $\tau_{r_{\rm max}}$ into the equation for optical depth distribution along the jet axis \citep{2019MNRAS.488..939P}: %
\begin{equation}
\label{eq:rcoreratio}
\frac{r_{\rm max}}{r_{\tau{\rm=1}}} = \tau_{r_{\rm max}}^{-1/(n+m(1.5+\alpha)-1)}.
\end{equation}
For a wide range of parameters, this ratio is $r_{\rm max}/r_{\tau=1} \approx 0.7$, and $r_{\rm max}$ differs from $r_{\tau=1}$ by $\approx 40\%$.
This implies that even if the core positions measured by Gaussian model fitting were unbiased, then the magnetic field estimated from the core shifts would be biased downwards by a factor $\approx1.3$ \citep{1998A&A...330...79L,2009MNRAS.400...26O} or even $\approx6$ \citep{2015MNRAS.451..927Z}.
This effect does not depend on the core shift itself and only scales the obtained magnetic field estimates uniformly by a constant factor. It could be easily compensated for by re-scaling the corresponding core shift estimates.

\section{Bias due to VLBI core Gaussian fitting}
\label{sec:biasgaussiancore}

\cite{1979ApJ...232...34B} model has a significantly asymmetric radial profile (Fig.~\ref{fig:bkslice}). To assess the bias due to its approximation with a Gaussian template, simulations with a realistic brightness distribution model are necessary.
Here we describe a study of the systematics in core shift measurements using brightness distribution of the Blandford-K\"{o}nigl jet model, constructing artificial visibility data sets and processing them with conventional methods.

In this section, we use the results of MOJAVE observations to construct and study the properties of our artificial sample.
We note that through the years MOJAVE group has constructed and published results for slightly different samples, including the MOJAVE-I sample \citep{2009AJ....137.3718L,2012A&A...545A.113P,2013A&A...558A.144C}, its extension in \citep{2017MNRAS.468.4992P}, as well as the  MOJAVE 1.5~Jy Quarter-Century sample \citep[1.5JyQC,][]{2019ApJ...874...43L}.
All of them are VLBI flux-density-limited samples of Doppler-boosted AGN jets. Their properties are comparable and do not differ significantly for our goals.

\subsection{Creation of the artificial sample}
\label{sec:sample_creation}

\begin{figure*}
  \centering
  \includegraphics[width=0.48\textwidth]{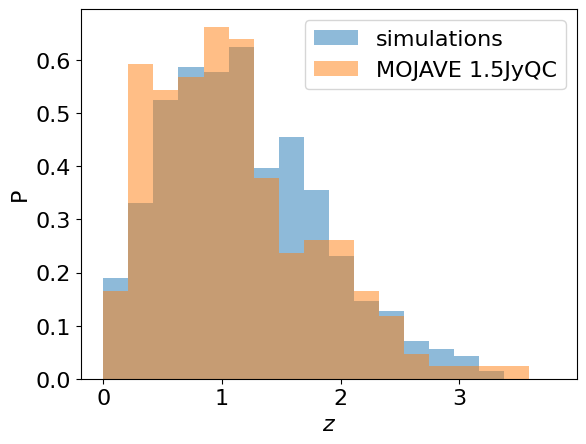}
  \includegraphics[width=0.48\textwidth]{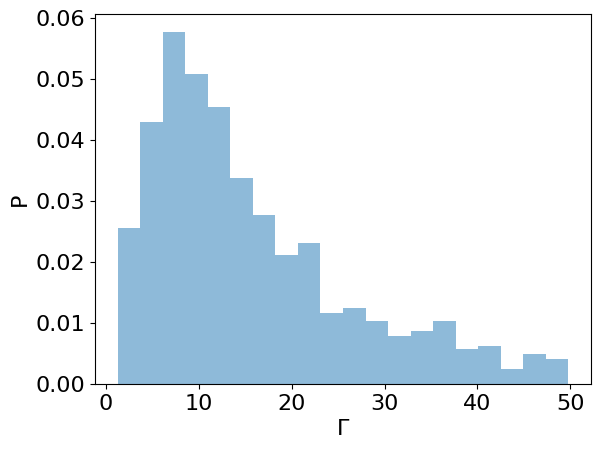}
      \caption{\textit{Left}: distribution of redshifts for the simulated sample (\textit{blue}) and sources with $z > 0.1$ from the MOJAVE 1.5JyQC sample (\textit{orange}). \textit{Right}: distribution of the Lorentz factors for the simulated sample.}
         \label{fig:zGamma}
  \end{figure*}

The brightness distribution of the \cite{1979ApJ...232...34B} model depends on several parameters. Only some combinations of these parameters are identifiable on the basis of the VLBI observations \citep{2019MNRAS.488..939P}. Using a uniform or random grid in the parameter space will result in sample parameter values, which are not typical for the sources observed in flux-density-limited samples (e.g. 1.5JyQC). Thus we construct a grid of model parameters conditioning them on the observed data in the following way. First, we draw a source with the parameters $(L, z, \theta, \Gamma)$ and the flux density $>1.5$~Jy from the intrinsic pure luminosity evolving luminosity function based on the 1.5JyQC sample. Here $L$~--- luminosity , $z$~--- redshift, $\theta$~--- viewing angle and $\Gamma$~--- bulk motion Lorentz factor. We draw $z$ from constant co-moving density between $z_{\rm min}=0.1$ and $z_{\rm max}=3.4$. Thus we obtained the power-law low-luminosity cut-off of the unbeamed luminosity function from \cite{2019ApJ...874...43L}. Then we draw $L$ from the obtained power-law distributed luminosity function and $\Gamma$ from the power-law with limits $\Gamma_{\rm min} = 1.25$ and $\Gamma_{\rm max} = 50$ and exponent $-1.4$, obtained from the Monte-Carlo modelling of the 1.5JyQC sample in \cite{2019ApJ...874...43L}. The viewing angle $\theta$ was drawn from the isotropical (i.e. uniform in $\cos{\theta}$) distribution. From the unbeamed luminosity and Doppler factor $\delta(\theta, \Gamma)$ we calculated the observed luminosity $L\delta^{p+\alpha}$, where we assume beaming exponent $p = 2$ for a continuous flow and spectral index $\alpha = 0$. If the observed flux corresponding to the parameters $(L, z, \theta, \Gamma)$ was larger than the threshold value 1.5~Jy, the source was accepted to the simulated sample. The distributions of redshifts and Lorentz factors of the simulated sources are presented in Fig.~\ref{fig:zGamma}.

As expected, the redshift distribution is close to that of the 1.5JyQC sample, while the distribution of $\Gamma$ corresponds to its best fit Monte-Carlo simulation  \citep[see fig.~11 in][]{2019ApJ...874...43L}. To obtain the value of the half-opening angle $\phi_{1/2}$, we used relation $\phi_{1/2} \approx \rho/\Gamma$ derived from  statistical modelling of the opening angles distribution  by \cite{2013A&A...558A.144C}, where $\rho$ is drawn from the normal distribution $N(\mu=0.13, \sigma=0.02)$. The resulted distribution of apparent opening angles $\phi_{\rm app}$ in the simulated sample is consistent with those obtained by \citet{2017MNRAS.468.4992P}: compare our median $\phi_{\rm app} = 20.3^{\circ}$ with their $\phi_{\rm app} = 21.5^{\circ}$.

We assumed a conical jet geometry with a radial velocity field and exponents of the magnetic field and particle density amplitude $m=1$ and $n=2$ (\autoref{sec:bkmodel}).
The first corresponds to a conical jet geometry with a constant bulk motion Lorentz factor $\Gamma$ and isothermal particle distribution. The second implies a dominant transverse magnetic field component. Together they result in a constant ratio of the emitting particles to the magnetic field energy densities. This choice is justified because most of the core shift measurements were used in deriving the physical properties of the jet under the equipartition assumption.
Note that the departures from the conical shape are seen close to the jet base up to the distance of $10^6$ gravitational radii \citep{2020MNRAS.495.3576K}. This affects mostly nearby sources with the given limited VLBI resolution.
We employed synchrotron radiation transfer coefficients for a tangled magnetic field model \citep{longair_1994} and assumed the optically thin spectral index $\alpha = 0.5$, which corresponds to $s=2.0$ in the particle energy spectra $N(\gamma) \propto \gamma^{-s}$. To limit the visible size of the jet, we included a steepening of the spectral index with the value 0.0001~pc$^{-1}$.
With this settings, the model spectrum is flat \citep{1979ApJ...232...34B}. We also assumed that the low- and high-frequency cut-offs in the spectrum fall outside of the chosen frequency range
due to an external jet scale and break in the energy distribution of emitting particles in the magnetic field \citep{1981ApJ...243..700K}.

To obtain the sample of the core shifts, we generated Stokes I images for frequencies 2.3, 8.1 and 15.4~GHz (S, X and U radio bands, respectively) numerically solving the radiative transfer equation using an adaptive \texttt{Dormand-Prince~5} type of Runge\-Kutta algorithm \citep{DORMAND198019}. As we assume radial velocity field, the Doppler factor $\delta$ changes from the near to the far side of the jet. To obtain the synthetic VLBI data sets, we substituted the observed visibilities of a single template data set with their predicted model values and added a Gaussian noise estimated from the scatter of the observed visibilities. We used real VLBI observations of the source 1458+718 from \cite{mojave6kinematicsanalysis} and \cite{2011A&A...532A..38S} as a template for ($u,v$)-coverage and noise estimation. To estimate the thermal noise from the observed visibilities, we employed the successive differences approach \citep{briggs}. We choose a source with a high declination to reduce possible effects due to an elongated VLBA beam. We used major axes of naturally weighted CLEAN beams for calculations: 0.81, 1.61 and 6.51 mas for 15.4, 8.1 and 2.3 GHz. As will follow in \autoref{sec:biasorigin}, for the straight jet model it is the core shift in units of the beam size that determines the value of the core shift bias. For elongated beams (i.e. sources with a low declination), we expect qualitatively the same behaviour with the effective resolution determined by the beam size along the core shift direction (\autoref{sec:compareplavin}).

Following the common practice, we modelled the resulting data sets in the ($u,v$)-plane at each frequency using simple Gaussian templates in \texttt{Difmap} \citep{1997ASPC..125...77S}. We automated the modelling procedure routine using several stopping and filtering criteria to choose the ``best'' number of components, thus mimicking the modelling process by a human. At each step, the procedure analyses the residual map and suggests new components estimating their parameters by \textit{the method of moments} \citep{Wasserman:2010:SCC:1965575}. We also employed the approach of \cite{2019MNRAS.485.1822P} of obtaining VLBI core parameters. It consists of estimating the extended emission by a list of CLEAN components outside of the core region and subtracting them from the data set in the ($u$, $v$)-plane. The obtained residuals are then modelled with a single Gaussian component. We estimated the true core shift value using the model brightness distribution and determined the observed shift of the VLBI core in the corresponding \texttt{Difmap} model fitted to the synthetic VLBI data. Note that our brightness distributions are perfectly aligned, i.e. the jet cone apex is always positioned in the phase center of the VLBI images in the model.

Finally our simulated sample includes 1000 sources. We filtered 22 sources with core component flux densities larger than 10~Jy and less than 100~mJy. We also excluded 3 outliers with the ratio of the observed to true core shifts between 8.1 and 15.4~GHz less than 0.2 and one source with the observed core shift between 8.1 and 15.4~GHz larger than 1~mas.

\subsection{Observed and true core shifts in the simulated sample}
\label{sec:coreshifts}

The median of the observed core shifts between 15.4 and 8.1 GHz for the MOJAVE-I sample is 0.114 mas \citep[see tab.~1 in ][]{2012A&A...545A.113P} for $z$ > 0.1.
This agrees with the value of our simulated sample of $0.114\pm 0.003$ mas for a circular Gaussian core model (hereafter, CG). Hereafter the error corresponds to the 95$\%$ confidence interval. The confidence interval for the median was obtained by applying the Binomial distribution \citep{conover}. This agreement is expected because we run our simulations assuming population properties inferred from the MOJAVE 1.5JyQC sample. The median core shift for the elliptical Gaussian core model (hereafter, EG) is $0.169\pm0.005$ mas.

\cite{2008A&A...483..759K} found a median core shift between 2.3 and 8.6 GHz as 0.44 mas for their sample of 29 sources observed in geodetic sessions with a rather high dispersion (see their fig.~3). In our simulated sample, the median observed core shift between 2.3 and 8.1 GHz is $0.537\pm0.013$ mas. This is close to the value 0.61 mas extrapolated from the typical core shift between 15.4 and 8.1 GHz, assuming that the core shift $\Delta r$ between frequencies $\nu_1$ and $\nu_2$ scales with $(\nu_1 - \nu_2)/(\nu_1 \nu_2)$ \citep{1998A&A...330...79L}. \cite{2019ApJ...871..143P} estimated the typical core shift between 2.3 and 8.6 GHz as 0.53 mas for a sample of 40 sources observed at 10 epochs or more, mainly during geodetic VLBI sessions. However, a different method of core parameters extraction was employed (\autoref{sec:sample_creation}) there. In our simulated sample, this method provides the median core shift 0.91 mas.
Note that both studies required specific criteria on the core shift measurement to be reliable, e.g. the same source structure at both bands \citep{2008A&A...483..759K}, prominent extended structure \citep{2019ApJ...871..143P}, etc. These criteria are potentially capable of lowering observed typical core shifts relative to core shifts estimated for our simulated sample.
\cite{2008A&A...483..759K} estimated the theoretical mean core shift from 0.13 to 0.30~mas for a flux-density-limited sample between 2.3 and 8.6~GHz for the \cite{1979ApJ...232...34B} model. The specific value depends on the distribution of the Lorentz factors and assumes a mean synchrotron luminosity $10^{44}$~erg/s and a mean redshift $z=1$. We note that this is significantly lower than in our simulated sample based on different underlying assumptions about the source population. This also could be the reason for a lower typical core shifts.

  \begin{figure*}
  \centering
  \includegraphics[width=0.48\textwidth]{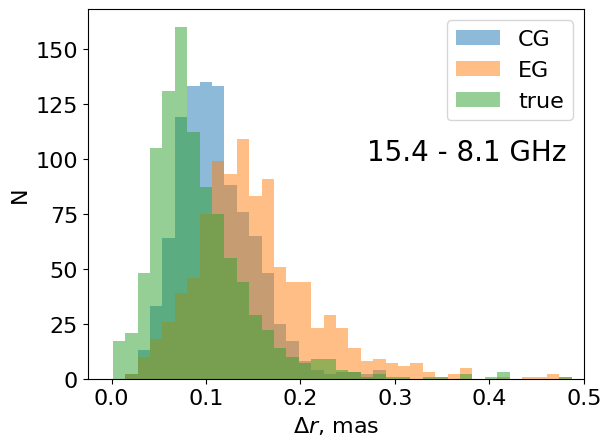}
  \includegraphics[width=0.48\textwidth]{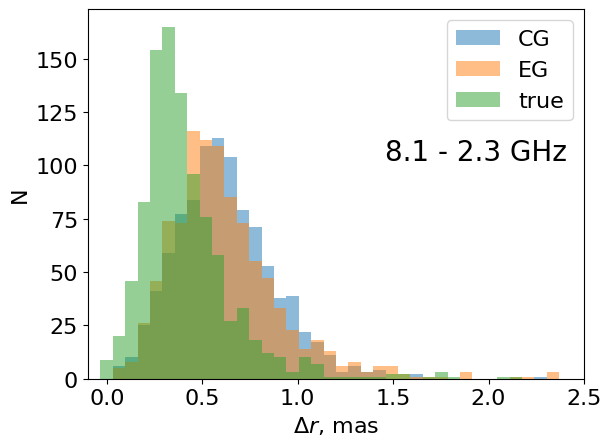}
      \caption{Distribution of the observed and true core shifts between 15.1 and 8.4 GHz (\textit{left}, 2 (true), 2 (CG), 8 (EG) points  are not shown) and between 8.1 and 2.3 GHz (\textit{right}, 1 (true), 2 (CG) and 14 (EG) points are not shown). Here CG means a circular Gaussian, and EG --- an elliptical Gaussian core model. The observed core shifts are measured with Circular Gaussian (CG, \textit{blue} colour) and Elliptical Gaussian (EG, \textit{orange} colour) core model.}
         \label{fig:shifts_xu_sx}
  \end{figure*}

In Fig.~\ref{fig:shifts_xu_sx} the observed and true core shifts obtained from the simulated sample are compared. The positive bias of the core shift estimates is evident --- the measured values are systematically larger than the true ones. The median of the true core shifts between 15.4 and 8.1 GHz is 0.09 mas,
while for 2.3 and 8.1 GHz it is
0.43 mas.

\cite{2012A&A...545A.113P} indicate a value of 0.05~mas as the random error of the core shift measurements between 15.4 and 8.1~GHz obtained using three approaches. This estimate includes a contribution from thermal noise, calibration error and mask size error. \cite{2019ApJ...871..143P} estimated the typical uncertainty of 8.6 - 2.3 GHz core shifts as 0.2 mas. The median bias of the estimated core shift between 8.1 and 15.4~GHz is 0.024 and 0.078~mas for circular and elliptical core models. For 8.6 - 2.3 GHz the corresponding median biases are 0.09 and 0.37~mas. These values are comparable to the random errors. However, even more important is the possible bias dependence on the source properties.

\subsection{Bias dependency on the resolution}
\label{sec:biasorigin}

  \begin{figure*}
  \centering
  \includegraphics[width=0.48\textwidth]{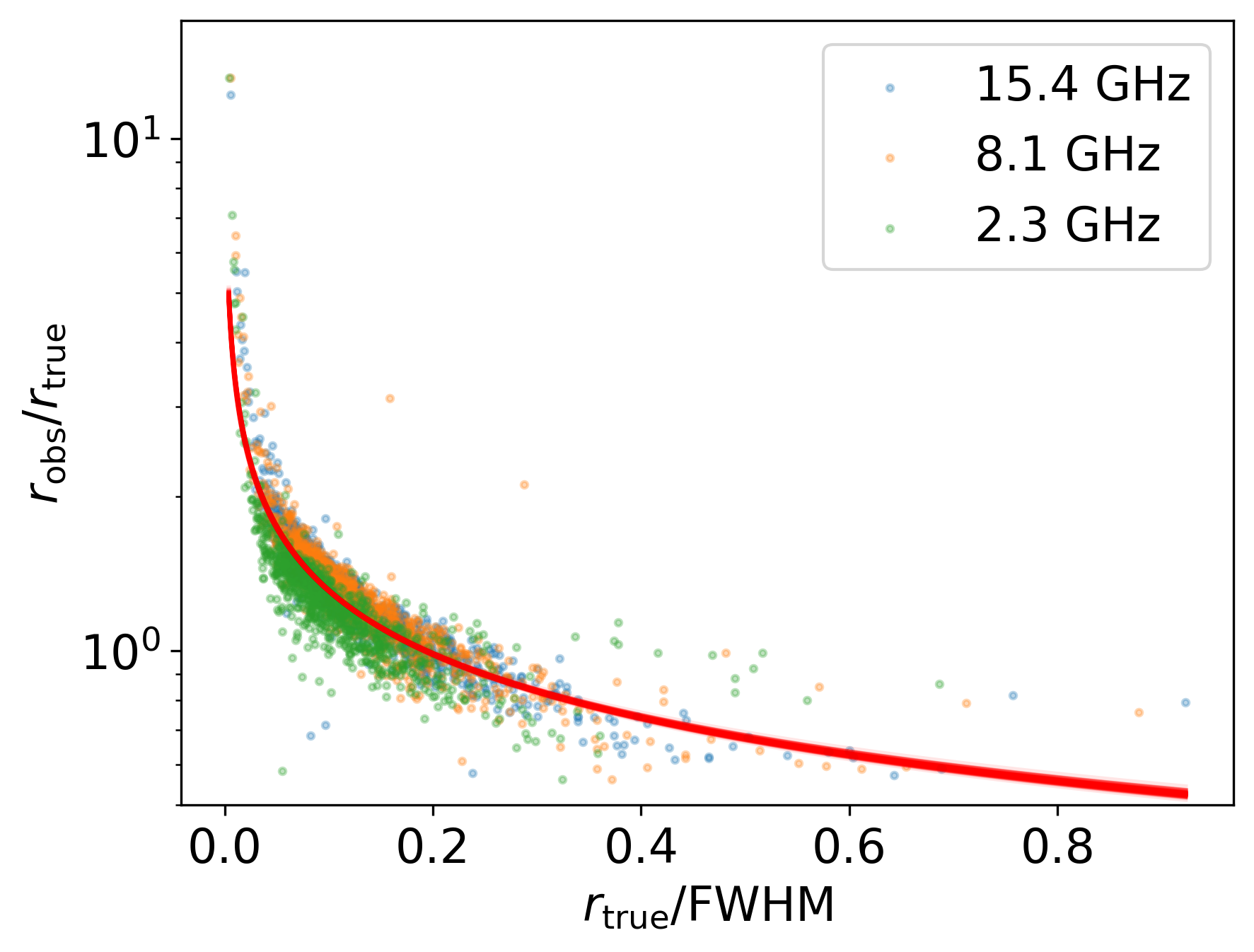}
  \includegraphics[width=0.48\textwidth]{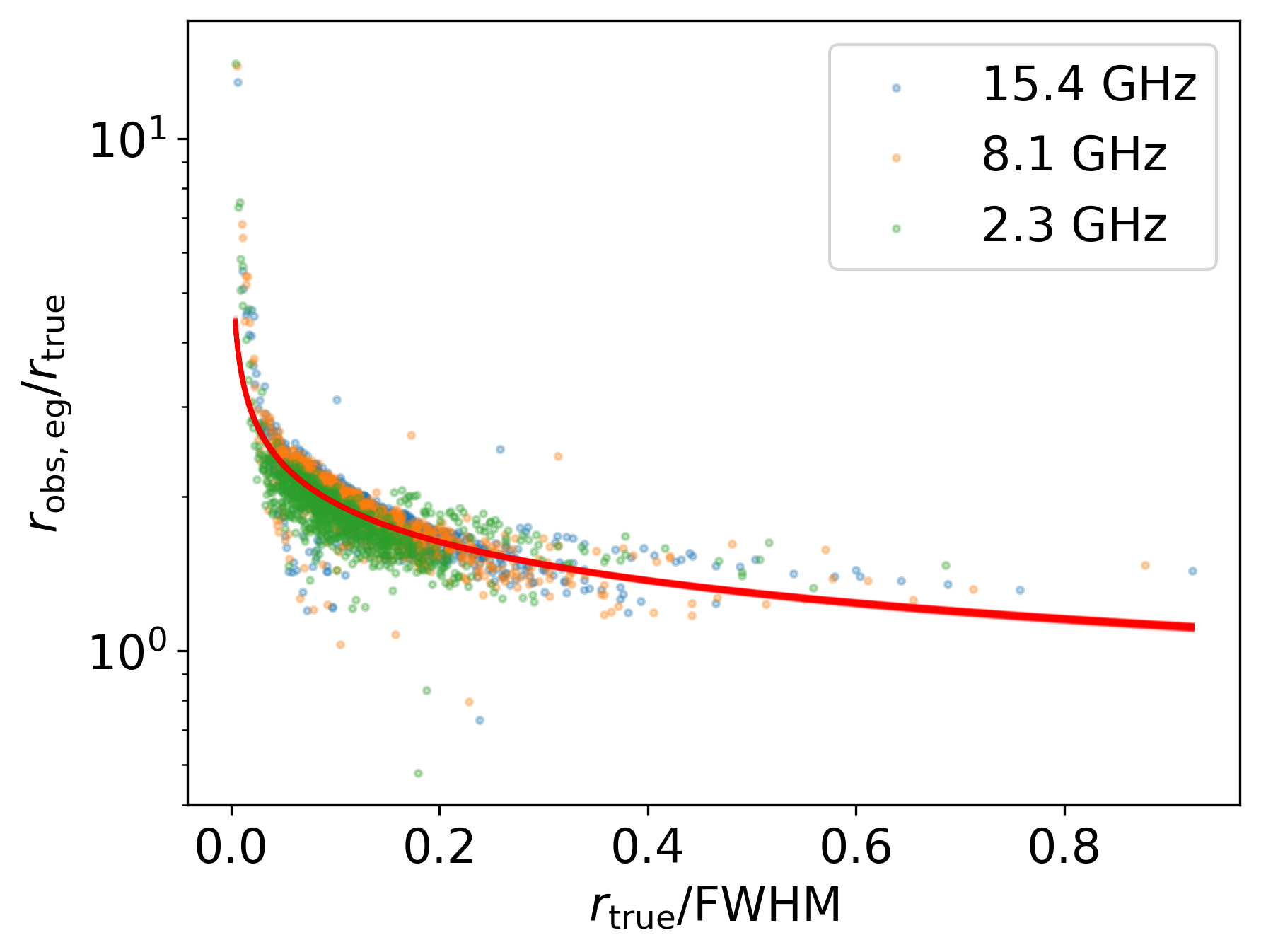}
      \caption{Dependence of the ratio between the observed $r_{\rm obs}$ and true $r_{\rm true}$ core projected distance on the ratio of the $r_{\rm true}$ to FWHM of the beam for circular (\textit{left}) and elliptical (\textit{right}) core templates. The points represent data at 15.4, 8.1 and 2.3 GHz (marked by \textit{blue}, \textit{orange} and \textit{green} colour). The thin red lines represent the samples from the posterior distribution of the parameters of the fitted model (\autoref{eq:rbias}). These lines are close and merging on this plot, thus the width of the red curve shows the spread of the model prediction.}
         \label{fig:r_bias}
  \end{figure*}

The core shift is determined by the difference in core positions at two frequencies. The dependence of the ratio of the observed $r_{\rm obs}$ to the true $r_{\rm true}$ projected core distance on the true core distance $r_{\rm true}$ expressed in units of the restoring beam FWHM for our simulated sample is shown in Fig.~\ref{fig:r_bias} for 15.4, 8.1 and 2.3 GHz.
Apparently, the core position is overestimated for the elliptical Gaussian core template. However, using circular Gaussians for core fitting may lead to an underestimation of $r_{\rm true}$. In our simulations, this occurs at high jet viewing angles when the core is modelled with two close components. The position of the first one (more compact with a flatter spectra, thus, treated as the core) lays upstream of the maximum of the model brightness.

\begin{figure*}
  \centering
  \includegraphics[width=0.48\textwidth]{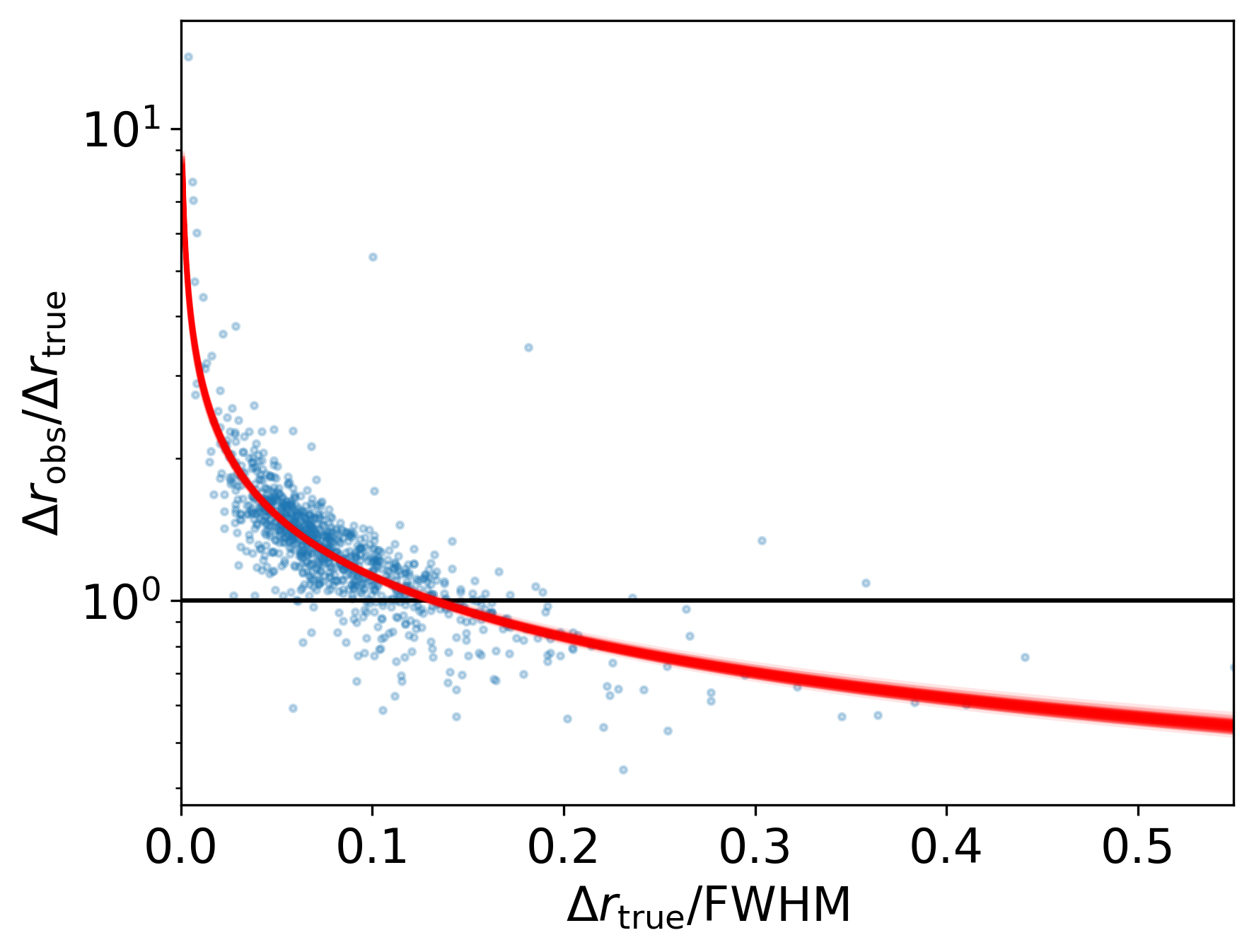}
  \includegraphics[width=0.48\textwidth]{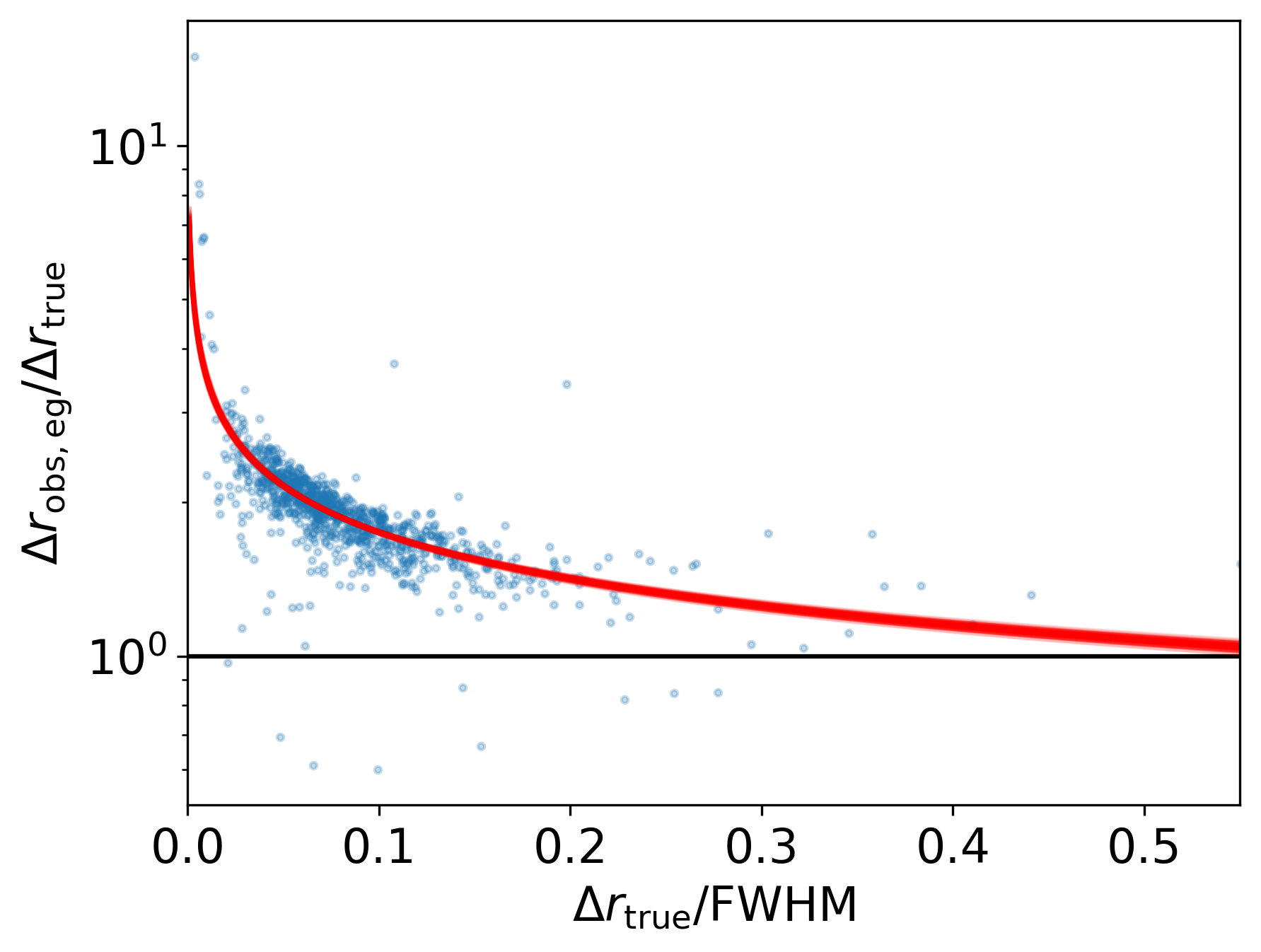}
      \caption{Dependence of the ratio between the observed $\Delta r_{\rm obs}$ and true $\Delta r_{\rm true}$ core shifts between 15.4 and 8.1 GHz on the ratio of the $\Delta r_{\rm true}$ to mean FWHM of the beams for circular (\textit{left}) and elliptical (\textit{right}) core templates. The thin red lines represent the samples from the posterior distribution of the parameters of the fitted model (\autoref{eq:dr_obs_to_dr_true_vs_drtrue}).
      Here these lines are close and merging, thus the width of the red curve shows the spread of the model prediction.}
         \label{fig:dr_bias}
  \end{figure*}

\begin{figure}
  \centering
  \includegraphics[width=\linewidth,trim=0cm 1cm 0cm 0cm]{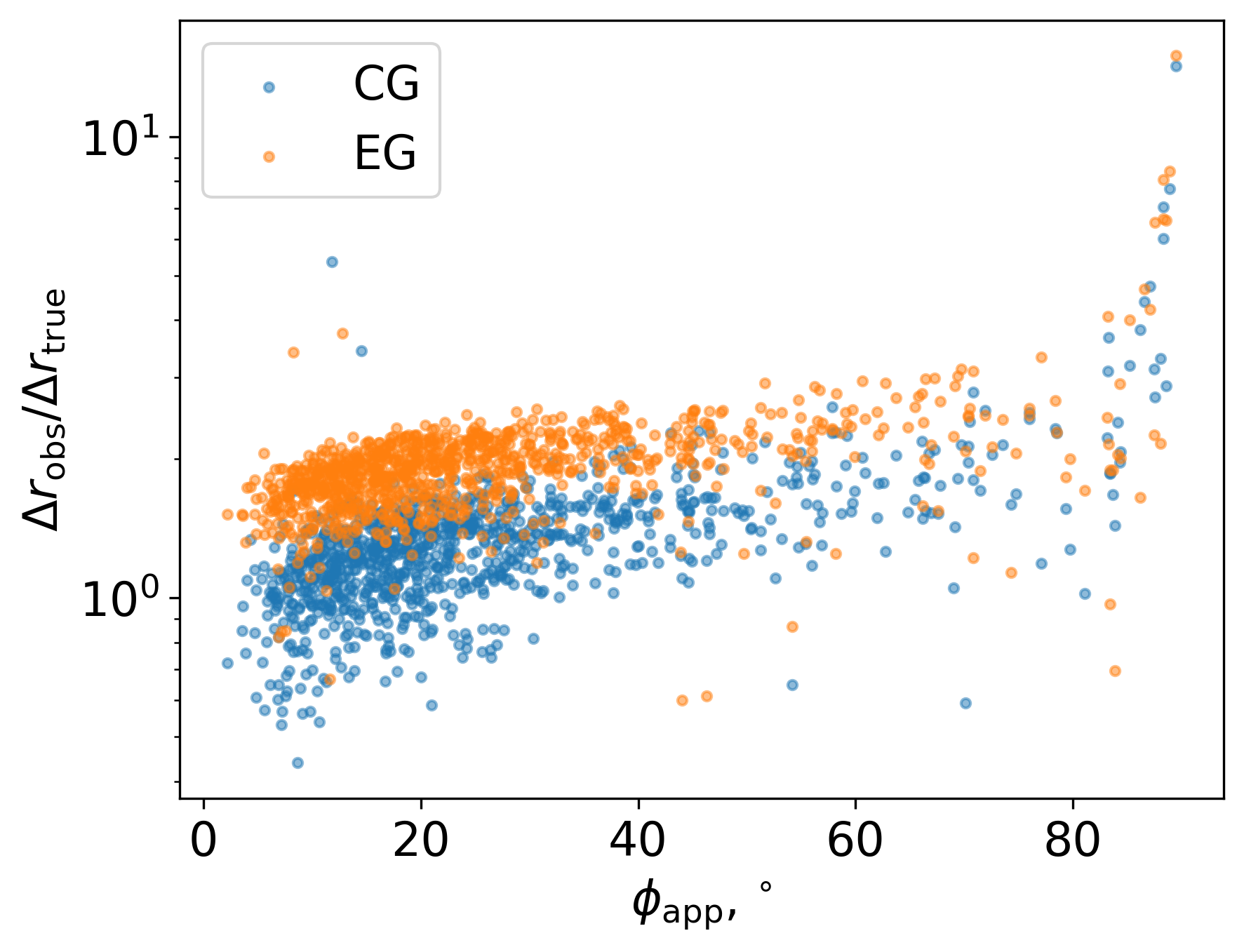}
      \caption{Dependence of the ratio between the observed and true core shifts between 15.4 and 8.1 GHz on the apparent jet opening angle for circular (\textit{blue}) and elliptical (\textit{orange}) core templates.}
         \label{fig:bias_phiapp}
  \end{figure}

We found that the relative angular resolution is the main factor which determines the fractional core position bias. This dependence in Fig.~\ref{fig:r_bias} can be well described by:
\begin{equation}
\label{eq:rbias}
    \frac{r_{\rm obs}}{r_{\rm true}} = C_0\left(\frac{r_{\rm true}}{\rm FWHM}\right)^{-k}
\end{equation}
%
with $C_0 = 0.507\pm0.008$ and $1.090\pm0.013$ for CG and EG.
Exponent $k = 0.41\pm0.02$ and $0.25\pm0.01$ for CG and EG. Note that the measurements at all frequencies are shown in Fig.~\ref{fig:r_bias}.
Assuming that the true position of the core for a given source is $r_{\rm true}(\nu) = C_1\nu^{-1/k_{\rm r}}$ and ${\rm FWHM}(\nu) = C_2\nu^{-1}$, the observed core position of this source at some frequency $\nu$ yields:
\begin{equation}
\label{eq:obsrcore}
        r_{\rm obs}(\nu) = C_0 C_1 \left(\frac{C_1}{C_2}\right)^{-k} \nu^{-1/k_{\rm r,eff}}
\end{equation}
and the ratio of the observed and true core shifts between two frequencies $\nu_1$ and $\nu_2$:
\begin{equation}
\label{eq:relative_cs_bias}
    \frac{\Delta r_{\rm obs}}{\Delta r_{\rm true}} = C_0\left(\frac{C_1}{C_2}\right)^{-k} \frac{\nu_1^{-1/k_{\rm r,eff}} - \nu_2^{-1/k_{\rm r,eff}}}{\nu_1^{-1/k_{\rm r}} - \nu_2^{-1/k_{\rm r}}}
\end{equation}
where the effective coefficient of the frequency dependence is:
\begin{equation}
\label{eq:k_r_eff}
    k_{\rm r,eff} = k_{\rm r}/(1+k(k_{\rm r}-1))
\end{equation}
equals to the true $k_{\rm r}$ only for $k_{\rm r} = 1$. In this case, the ratio of the observed to the true core shift is frequency independent. As $\Delta r_{\rm true} \propto C_1$ and $\mathrm{FWHM} \propto C_2$, this case also can be written via the true core shift in units of the mean restoring beam FWHM:
\begin{equation}
\label{eq:dr_obs_to_dr_true_vs_drtrue}
    \frac{\Delta r_{\rm obs}}{\Delta r_{\rm true}} = C_{\Delta r}\left(\frac{\Delta r_{\rm true}}{\rm    \langle FWHM \rangle}\right)^{-k}
\end{equation}
where
\begin{equation}
\label{eq:Ctau}
    C_{\Delta r} = \left(\frac{\nu_1 + \nu_2}{2(\nu_1-\nu_2)}\right)^{-k} C_0
\end{equation}
Here $\langle \mathrm{FWHM} \rangle$ --- mean size of the beam for two frequencies and $C_0$ --- constant from \autoref{eq:rbias}. For 15.4 and 8.1 GHz we obtain $C_{\Delta r} = 0.41$ and for 8.1 and 2.3 GHz $C_{\Delta r} = 0.53$. This relation should hold at any pair of frequencies provided that the true core shift and the resolution scales linearly with the wavelength.
The corresponding dependence is shown in Fig.~\ref{fig:dr_bias} for the core shift between 15.4 and 8.1 GHz with its fit by \autoref{eq:dr_obs_to_dr_true_vs_drtrue}, which gives $C_{\Delta r} = 0.42\pm0.02$ and $k = 0.43\pm0.02$ for CG. This agrees with our estimates of $C_{\Delta r}$ from \autoref{eq:Ctau} and fits of $C_0$ from \autoref{eq:rbias}. For EG, we obtained $C_{\Delta r} = 0.87\pm0.03$ and $k = 0.30\pm0.02$. For both CG and EG core templates, the overestimation of the core shift is the most pronounced for sources with the largest apparent opening angles $\phi_{\rm app}$ (Fig.~\ref{fig:bias_phiapp}).
In our artificial sample, these are the sources with the smallest jet viewing angles, i.e.\ blazars. Radio galaxies are observed at larger viewing angles, have smaller $\phi_{\rm app}$ \citep{2017MNRAS.468.4992P} and, consequently, practically unbiased core shifts.

For the core shift between 8.1 and 2.3 GHz, the corresponding values are $k = 0.34\pm0.02$ and $C_{\Delta r} = 0.57\pm0.02$ for CG. This is slightly different than estimated from \autoref{eq:rbias} and \autoref{eq:Ctau}. It could be due to a different VLBI array configuration at 2.3 GHz relative to other bands used as a template in our simulations.

\subsection{Comparison with existing observational estimates of the bias}
\label{sec:compareplavin}

A recent extensive study of the core shift effect variability \citep{2019MNRAS.485.1822P} analysed possible systematic biases as well. However, there are several differences in the analysis and the assumptions with our work. First, they focused on 2.3 and 8.6 GHz data obtained from different configurations of VLBI arrays. However, more importantly, is that they have assumed that the bias in core shift measurements is proportional to the beam FWHM averaged between the two frequencies, whereas we chose the dependence that fits simulated data well. Moreover, they employed a different approach to the estimation of the core parameters.

\cite{2019MNRAS.485.1822P} estimated the proportionality coefficient between core shift bias and mean beam FWHM to be $0.14$. Assuming that the effect at each frequency is proportional to its beam size, this implies that the bias of the core position is
$0.10 \cdot \langle {\rm FWHM} \rangle$. Applying the same approach to the core parameter estimation as in \cite{2019MNRAS.485.1822P} to our simulated sample, we
obtain the median bias of $0.11 \cdot \langle {\rm FWHM} \rangle$.
Note that there are other possible sources of uncertainties \citep[see][]{2019MNRAS.485.1822P}, which are out of the scope of this work. However, if the close agreement does not occur by chance, it suggests that the remaining measurement uncertainties only increase the statistical variance and do not make the results more biased.
It also implies that significant beam elongation present in \cite{2019MNRAS.485.1822P} does not significantly change the effect.

The assumption adopted in \citet{2019MNRAS.485.1822P} leads to different dependence of the fractional core shift bias on the true core shift between 2.3 and 8.1 GHz:
\begin{equation}
\label{eq:plavin}
    \frac{\Delta r_{\rm obs}}{\Delta r_{\rm true}} = 1 + 0.10\left(\frac{\Delta r_{\rm true}}{\langle{\rm FWHM}\rangle}\right)^{-1}.
\end{equation}
This is more consistent with our results obtained for the elliptical Gaussian core model (Fig.~\ref{fig:dr_bias}, left), however, with a steeper dependence on the true core shift $-1$ vs.\ $-0.25$.

Finally, we also compared the different procedures of the core parameter estimation mentioned in \autoref{sec:bkmodel}. It occurs that if a circular Gaussian is used to model the core, then the subtraction-based approach of \cite{2019MNRAS.485.1822P} almost always overestimates core shifts compared to the traditional approach of a modelling structure with a set of Gaussians. For 2.3--8.1~GHz, the median ratio and 95\,\% inter-quantile ranges are 1.64 and (0.99, 3.04), while for 8.1--15.4~GHz, they are 1.59 and (0.26, 3.46). However, using an elliptical Gaussian as a core model leads to comparable results with the median ratio of 1.07 and 95\,\% inter-quantile range (0.67, 1.99). We found that the core flux density is also overestimated~--- the median ratio of about 1.25 for all bands with interval (0.8, 2.2). At the same time, the most biased observable for subtraction-based core modelling is the core size. It overestimates true core sizes up to an order of magnitude, while the traditional approach estimates the true core size within a factor of 2.
Here we calculated the true core size as half-width of the model jet at the position of the observed core. For all observables, the exact value of the overestimation depends on the core flux density fraction, with core dominant jets providing more consistent results. We interpret this as a result of modelling the extended structure by a set of CLEAN components prior to fitting a single Gaussian to the core region. CLEAN components, which are $\delta$-functions, always contribute to a correlated flux density at the longest baselines which could significantly influence the subsequent estimation of the core parameters.

\section{Method to compensate for the bias}
\label{sec:how2compensate}

\begin{figure*}
  \centering
  \includegraphics[width=0.48\textwidth]{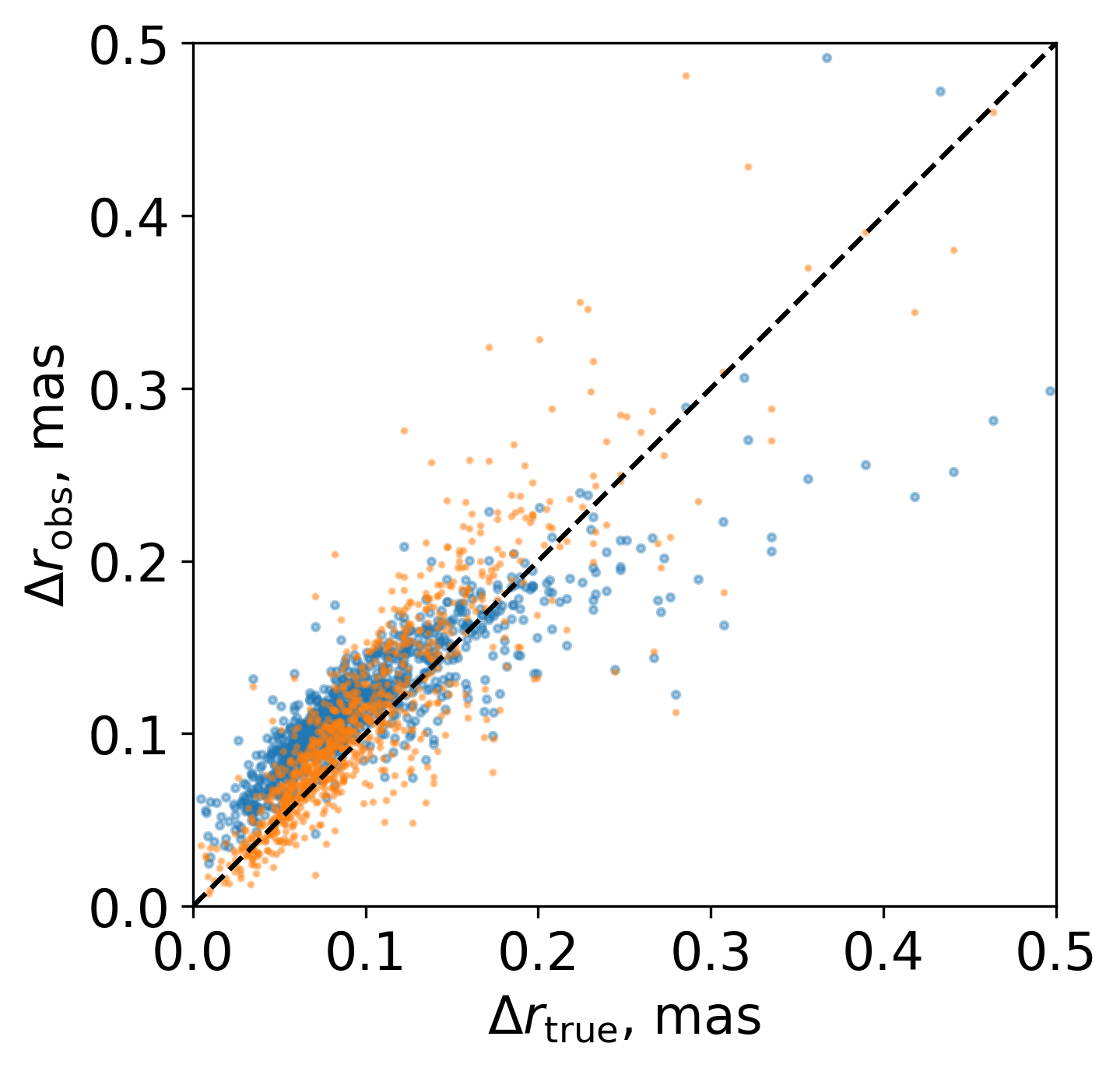}
  \includegraphics[width=0.48\textwidth]{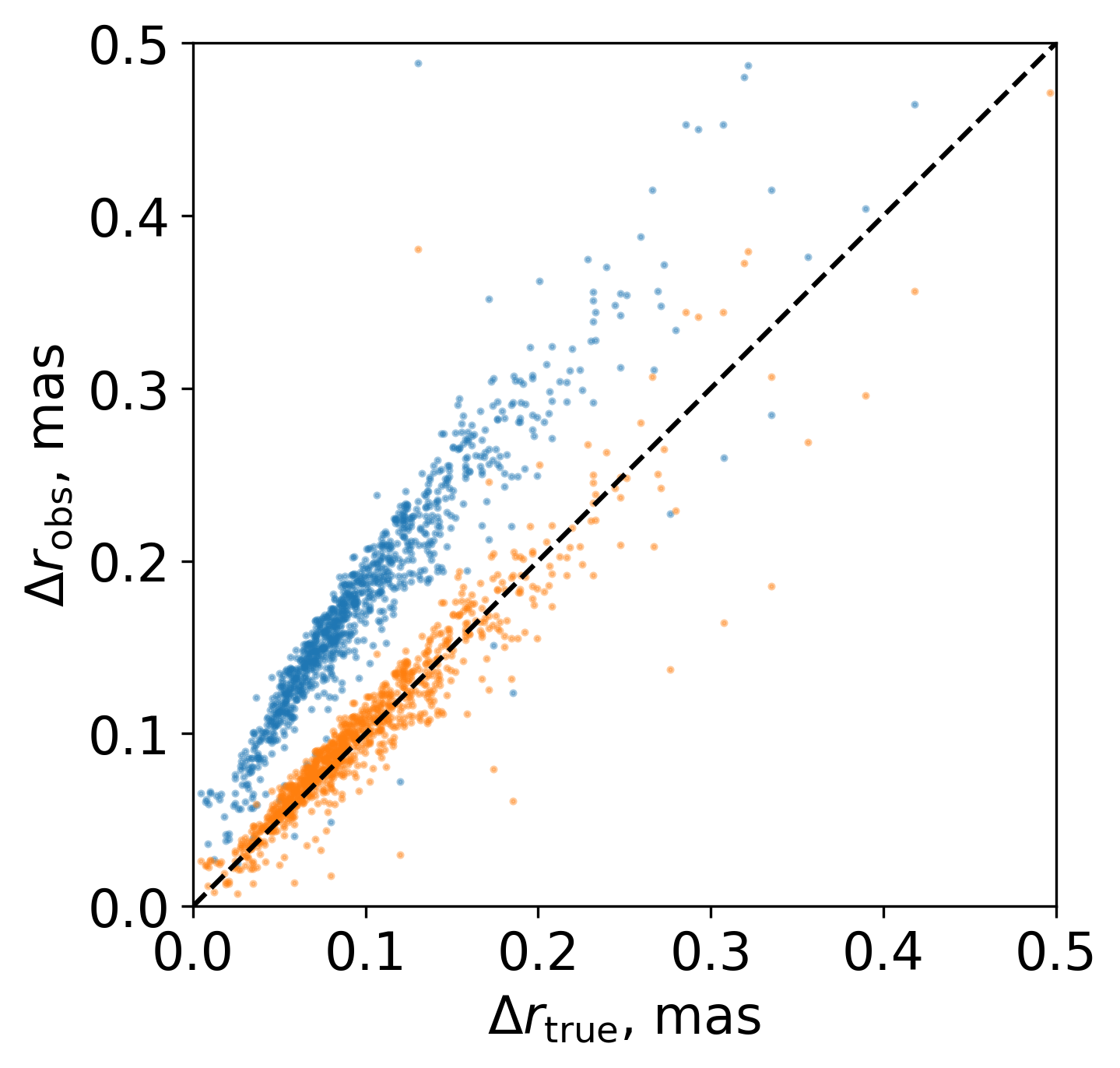}
      \caption{Raw observed (blue) and corrected (orange) core shifts vs true core shifts between 15.4 and 8.1~GHz for circular (\textit{left}) and elliptical Gaussian (\textit{right}) according to \autoref{eq:compensating_dr_bias}. The dashed line represents the equality between the observed and true values.}
         \label{fig:correction}
  \end{figure*}

\begin{figure}
  \centering
  \includegraphics[width=0.99\columnwidth]{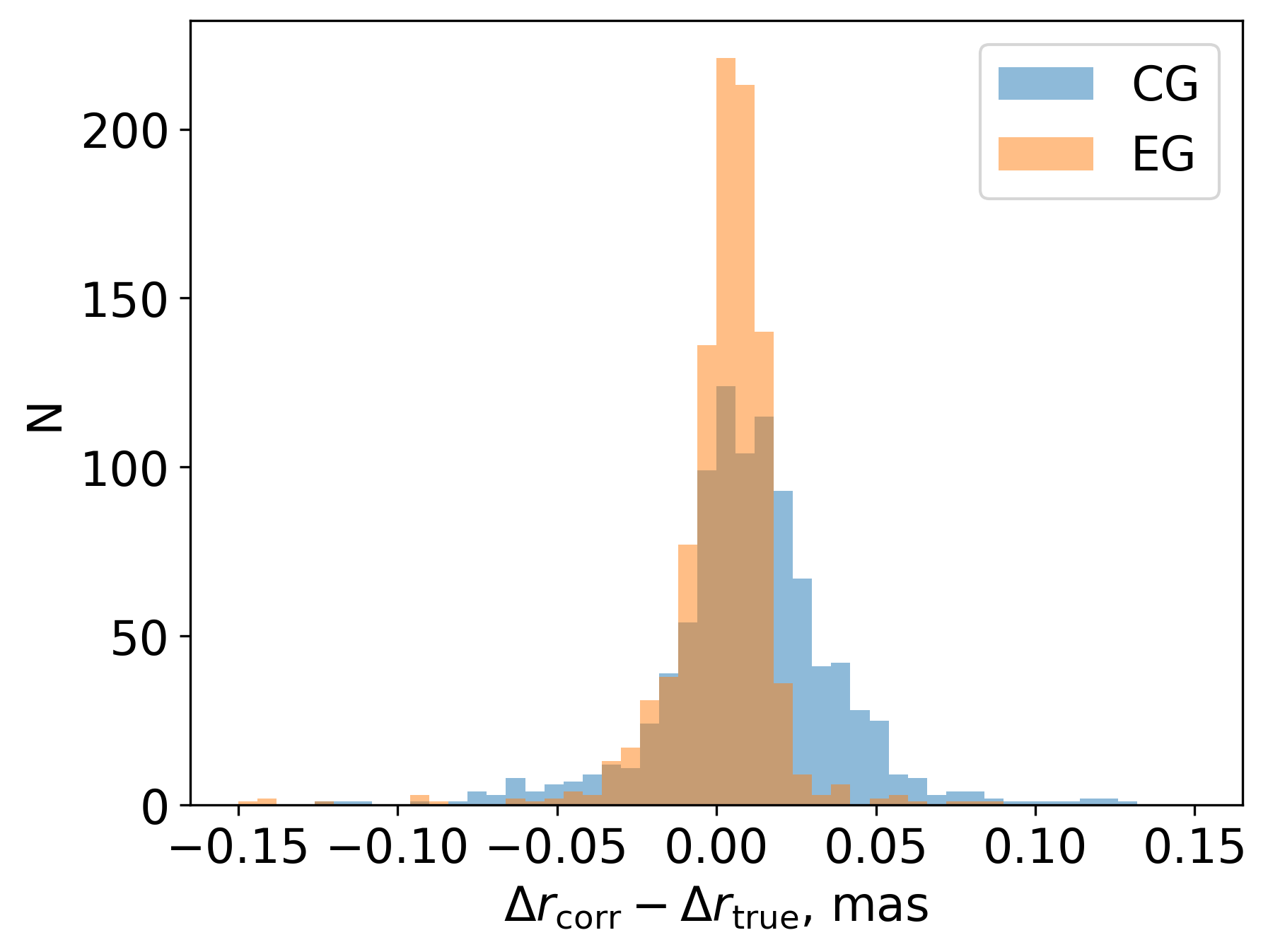}
      \caption{Residuals between the true and measured but bias-corrected core shifts between 15.4 and 8.1~GHz for the circular (\textit{blue}) and elliptical Gaussian core models (\textit{orange}).}
         \label{fig:correction-residual}
  \end{figure}

If only the measurements of the core shift between two frequencies $\nu_1$ and $\nu_2$ are available, one can assume $k_{\rm r}$ = 1 and obtain the relation for compensating the bias from \autoref{eq:dr_obs_to_dr_true_vs_drtrue} and \autoref{eq:Ctau}:
\begin{equation}
\label{eq:compensating_dr_bias}
    \frac{\Delta r_{\rm true}}{\langle {\rm FWHM} \rangle} = \left(\frac{\nu_1 + \nu_2}{2(\nu_1-\nu_2)}\right)^{k/(1-k)} C_0^{1/(k-1)} \left(\frac{\Delta r_{\rm obs}}{\langle {\rm FWHM} \rangle} \right)^{1/(1-k)}.
\end{equation}
Here constants $C_0$ and $k$ are estimated in \autoref{sec:biasorigin} for both circular and elliptical core models.
Note that \autoref{eq:compensating_dr_bias} can be employed for any pair of frequencies provided both true core shift and the resolution scale linearly with the wavelength.
In Fig.~\ref{fig:correction}, we plot the measured and true core shifts between 15.4 and 8.1 GHz together with those corrected for the bias using \autoref{eq:compensating_dr_bias}. Here the residuals between the corrected and true values correspond to $\sigma_{\rm MAD}=0.02$ and 0.01 mas for circular and elliptical Gaussians (Fig.~\ref{fig:correction-residual}), where $\sigma_{\rm MAD}$ is the robust estimate of standard deviation \citep{LEYS2013764}. The biases of both compensations are 0.010 and 0.005 mas for CG and EG, respectively.

If observations at more than two frequencies are analysed, then \autoref{eq:obsrcore} can be fitted to the measured core shift frequency dependence. Here $C_0$ and $k$ are estimated in this work (see \autoref{sec:biasorigin}) and $C_2$ can be estimated from fitting the inverse frequency dependence to beam sizes. After $k_{\rm r, eff}$ and $C_1$ are found, the can then be used to estimate true core shifts using the original relation $r_{\rm true}(\nu) = C_1 \nu^{-1/k_{\rm r}}$.

Finally, we warn readers that the estimates of $k$ and $C_0$ (thus $C_{\Delta r}$) could depend on the VLBI array configuration, as shown in \autoref{sec:biasorigin}. This could introduce uncertainty in the core shift correction procedure described here. Thus the most rigorous way to compensate for the core shift bias for a particular VLBI array configuration would be to estimate coefficients $k$ and $C_{\Delta r}$ performing a series of simulations using a jet model \citep[e.g.][]{1979ApJ...232...34B} and a specific ($u$, $v$)-coverage.

\section{Implications}
\label{sec:astrophysics}

\subsection{Equipartition magnetic field estimates}
\label{sec:eqbfields}
The measured shifts are widely used in estimating the reference values (at 1~pc from the apex) of
the magnetic field $B_{\mathrm{1}}$ and particle density $K_{\mathrm{1}}$  assuming equipartition \citep{1998A&A...330...79L, 2005ApJ...619...73H,2009MNRAS.400...26O,2012A&A...545A.113P}:
\begin{equation}
\label{eq:bdr}
    B_\mathrm{1}^{\rm eq} \propto (\triangle r_\mathrm{obs})^{0.75k_\mathrm{r}}.
\end{equation}
Here we should account for both biases: one is due to defining the core shift as the $\tau=1$ region but measuring using the position of the Gaussian core (\autoref{sec:tauoneorrmax}), and the other is due to the measured position of the Gaussian core not representing the maximum intensity (\autoref{sec:biasgaussiancore}). They affect $\triangle r$ in different directions and compensate each other for $\triangle r_{\rm true}/ {\rm FWHM}$ = 0.1 for CG and 0.4 for EG according to Fig.~\ref{fig:dr_bias} and fits with  \autoref{eq:dr_obs_to_dr_true_vs_drtrue}. Note that the magnetic field value $B_{\rm core}$ at the core itself turns out less biased. The distance of the core from the jet apex $r_{\rm core} \propto \triangle r_\mathrm{obs}$ for $k_{\rm r}$ = 1 \citep{1998A&A...330...79L}. Thus the magnetic field in the core weakly depends on the measured core shift as $B_{\rm core}^{\rm eq} = B_1^{\rm eq} r_{\rm core}^{-1} \propto (\triangle r_\mathrm{obs})^{-1/4}$.

\cite{2009MNRAS.400...26O} have estimated the 1$\sigma$ uncertainty of the magnetic field strength changing from 15$\%$ at 15.4 GHz to 100$\%$ and larger at 5 GHz, assuming equipartition and only considering the errors in the core shift measurements. More robust estimates should include others sources of errors
as those of Doppler factors, jet opening and viewing angles, and spectral indices.
Thus, the biases considered here should be treated as a lower limit of the accuracy of the magnetic field estimation. However, as averaging across multiple observations does not decrease a bias, it has certain astrophysical implications discussed below.

\cite{2014ApJ...796L...5N} found that magnetic fields inferred from \autoref{eq:bdr} are smaller than the $B_1$ estimates from the modelling of the spectral energy distribution (SED) by a factor of 2. This result is based on a sample of blazars from \cite{2014Natur.510..126Z} with a core shift estimated by \cite{2012A&A...545A.113P}. These authors propose that jets could have a non-uniform distribution of the magnetic field across the jet. An alternative explanation is that magnetic fields inferred from the core shift measurements are overestimated. \cite{2014ApJ...796L...5N} associate this with radio cores not being photospheres due to the synchrotron self-absorbtion (SSA) process, but due to a break in a lower energy cutoff of the energy distribution of the emitting particles. In our simulations, the sources with the smallest viewing angles are the most biased in terms of the core shift. This could reconcile the lower measured $B_1$ with the corresponding SED estimates within the SSA model of radio cores.

\cite{2014Natur.510..126Z} found observational evidence for the magnetically arrested disc (MAD) \citep{1974Ap&SS..28...45B,2003PASJ...55L..69N} scenario in which magnetic fluxes of the jet $\Phi_\textrm{j} \propto B_{\rm p} R_{\rm j}^2$ should be comparable to magnetic fluxes $\Phi_\textrm{MAD}$
threading a black hole. Here $B_{\rm p}$ is the poloidal magnetic field in the jet at some distance where the jet radius equals to $R_{\rm j}$. As $\Phi_\textrm{j}$ is
estimated using magnetic field amplitudes (\autoref{eq:bdr}), it is overestimated for blazars. Another assumption lowering $\Phi_\textrm{j}$ was that $\Gamma \theta \propto 1$.
\citet{2005AJ....130.1418J,pushkarev2009,2013A&A...558A.144C} estimated  $\Gamma \theta \sim 0.1$ in the optically thin part of the jets. Accounting for this and for the biased core shift estimates results in the decrease of $\Phi_\textrm{j}$ on the order of the magnitude for blazars \citep[see also ][]{Finke_2019}.


\subsection{Non-equipartition magnetic field estimates}
\label{sec:noeqbfields}

A magnetic field can be also estimated from the core shift without equipartition assumptions using equation (8) from \cite{2015MNRAS.451..927Z}:
\begin{equation}
\label{eq:noeq}
    B_{\rm noeq}(r)\simeq{3.35\times 10^{-11} D_L\,\delta\, (\triangle r_\mathrm{obs})^5\tan^2\phi\over r(\nu_1^{-1}-\nu_2^{-1})^5\left[(1+z)\sin \theta\right]^3 F_\nu^2}
\end{equation}
where $r$ is distance in pc, $F_{\nu}$ --- flux density of an inner jet integrated over both its optically thick and thin parts in Jy, $D_L$ --- luminosity distance in Mpc, $\nu_1$ $<$ $\nu_2$ --- frequencies of observations in GHz, and the numerical coefficient is calculated for the exponent of the emitting particles energy distribution $s=2$. Strong dependence on $\triangle r_\mathrm{obs}$ implies that the found systematic error of the core shift could bias the magnetic field estimates up to an order of magnitude.

\cite{2015MNRAS.451..927Z} estimated magnetic fields from core shift measurements using both equipartition (\autoref{eq:bdr}) and a more general approach (\autoref{eq:noeq}). They found $\langle B_{\rm noeq}/B_{\rm eq}\rangle$ = 1.6 for blazars and $\langle B_{\rm noeq}/B_{\rm eq}\rangle$ = 0.09 for radio galaxies. This ratio is modified in the presence of the core shift bias as $\langle B_{\rm noeq}/B_{\rm eq}\rangle = (\triangle r_{\rm obs}/\triangle r_{\rm true})^{5-0.75}\times\langle B^{\rm unbiased}_{\rm noeq}/B^{\rm unbiased}_{\rm eq}\rangle$. Here $B_{\rm unbiased}$ --- magnetic field estimate which would be obtained if core shift measurements were unbiased.
From our sample, the mean of $\triangle r_{\rm obs}/\triangle r_{\rm true}$
for sources with $\theta > 20^\circ$ (i.e. radio galaxies), corrected for bias estimated in \autoref{sec:tauoneorrmax}, is 0.73. Then, for blazars to have the same ratio of the unbiased estimates $\langle B^{\rm unbiased}_{\rm noeq}/B^{\rm unbiased}_{\rm eq}\rangle$, the mean ratio $\langle \triangle r_{\rm obs}/\triangle r_{\rm true} \rangle$ should be 1.44. In our simulated sample, this is achieved for sources with $\theta < 1.4^\circ$.
\cite{2015MNRAS.451..927Z} attribute the effect of different $\langle B_{\rm noeq}/B_{\rm eq}\rangle$ for blazars and radio galaxies to different viewing angles $\theta$ and the corresponding difference in relativistic beaming. However, it can also be explained by the systematics of the core shift measurements.

\cite{2017MNRAS.465.3506P} found that jet power $P_{\rm j}$ values computed from the core shift estimated according to \cite{2015MNRAS.451..927Z} are an order of magnitude lower than those estimated through the extended radio lobe luminosity. As the jet power $P_{\rm j} \propto B^{2}$ and $B \propto \triangle r^{5}$, this disagreement could be explained if the measured core shifts are overestimated by a factor of 1.26, consistent with our findings.

\subsection{AGN jets: flares, speeds and geometry}
\label{sec:flares}

\cite{2019MNRAS.485.1822P} found that core shifts vary significantly during flares. Their results imply that flares are due to an increase in the emitting particle density and a decrease of the magnetic field.
As follows from the flare model presented in \cite{2019MNRAS.485.1822P}, see their fig.~13, when a flare reaches the core at some frequency, it shifts the core downstream. This increases the core separation from the jet base and, according to \autoref{sec:coreshifts}, decreases its fractional bias. Thus the quiescent state has the most fractionally biased core shift estimate. Contrary to expectations, it could be that the most accurate core shift measurements in terms of a fractional bias are possible during the flaring state. Not accounting for this effect could potentially underestimate departures from the equipartition in cores during flares, as found in \cite{2019MNRAS.485.1822P}.

\cite{2017ApJ...834...65A} found evidence for non-conical jet shapes in radio-loud AGN jets in the core region using core sizes $R$ measured at at least four frequencies. The distances of radio cores from the central engines were estimated using measured core shifts from \cite{2012A&A...545A.113P} and assuming $r_{\rm core} \propto \triangle r_{\rm obs}/\nu$ (i.e. $k_{\rm r}$ = 1). For 56 sources with fit statistic $R^2 < 0.85$, the obtained outflow geometry is characterised by the coefficient $\epsilon$ of the radial core size dependence $R \propto r^{\epsilon}$, which is consistent with a quasi-parabolic outflow with mean $\langle \epsilon \rangle$ = 0.87. From \autoref{eq:dr_obs_to_dr_true_vs_drtrue}, the true core shift is related to the observed one as $\Delta r_{\rm true} \propto  (\Delta r_{\rm obs})^{1/(1-k)}$. Then, in the presence of a bias, the underlying dependence $R \propto r^{\epsilon} \propto \Delta r_{\rm true}^{\epsilon}$ transforms to $R \propto ( \Delta r_{\rm obs} )^{\epsilon/(1-k)}$, changing the effective exponent to $\epsilon^{\rm eff} = \epsilon/(1-k)$. In \autoref{sec:biasorigin}, we estimated $k \approx 0.42$ in the case of fitting the core with a circular Gaussian template. Thus the exponent $\langle \epsilon^{\rm eff} \rangle = 0.87$ measured by \cite{2017ApJ...834...65A} corresponds to the intrinsic $\langle \epsilon \rangle = 0.50$, implying parabolic jet shape. However, \cite{2017ApJ...834...65A} also employed the elliptical core model in their estimation of $\langle \epsilon \rangle$. According to \autoref{sec:biasorigin}, modelling the core with an elliptical Gaussian results in a lower $k$ and, consequently, higher $\langle \epsilon \rangle$. Thus our estimate of the intrinsic $\langle \epsilon \rangle$ should be considered as a lower bound.

The overestimation of $\epsilon$ due to the core shift bias could be interpreted as follows.
In the parabolic accelerating outflow, the magnetic field and the particle density gradients are lower than those in a conical constant speed jet. This leads to a steeper than $\nu^{-1}$ frequency dependence of the core shift for distances closer to the jet base (i.e. higher frequencies), as demonstrated by \cite{2011ApJ...737...42P}. Since resolution scales linearly with wavelength, and the bias is determined by the ratio of the true core shift to the resolution (\autoref{sec:biasorigin}), core shift estimates at higher frequencies will have a larger fractional bias, effectively increasing the measured $\epsilon$. However, this assumes that the model brightness distribution is close to that of \cite{1979ApJ...232...34B} for a conical jet used in our simulations. In fact, \cite{2018A&A...614A..74A} found more elongated cores at 43 GHz relative to 15 GHz for a fraction of sources in their complete S5 polar cap sample \citep{1986A&A...168...17E}. This also could be a manifestation of a radio core at 43~GHz residing in the parabolic accelerating part of the jet.

If a source demonstrates flaring activity, it is possible to derive the kinematics of the jet innermost regions based on the measured core shift and multi-frequency time delays \citep{2011MNRAS.415.1631K,2014MNRAS.437.3396K,2018MNRAS.475.4994K,2019MNRAS.486..430K}.
Assuming that the peak of a flare at a given frequency is observed when a moving jet feature passes through the region of the core at that frequency, it is possible to estimate the apparent projected speed of the component by comparing the light curve time delay, $\triangle t$, with the core shift for a pair of frequencies, $\triangle r$: $\mu_{\textrm{app}}$ = $\triangle r$~/~$\triangle t$. \cite{2014MNRAS.437.3396K} obtained an estimate of the apparent angular velocity $\mu_{\textrm{app}}$ = 0.7 mas $\textrm{yr}^{-1}$, which is 2-8 times larger than the speed estimated using VLBI kinematic data \citep{2013AJ....146..120L,2010ApJ...715..362J}. These papers propose several explanations for this discrepancy, e.g. truly fast components or VLBI not measuring true plasma speed. Systematically overestimated core shifts could at least partially explain this controversy. This especially holds for blazars, inclined at small angles to the line of sight and demonstrating the largest fractional bias.

\subsection{Astrometric implications}
\label{astrometry}

Precise positions of AGNs are measured with VLBI using either single-band phase delays or multi-band group delays of the arriving waves. The effects of the source structure and its frequency dependence on these positions were extensively studied: see, e.g. \cite{2009A&A...505L...1P}, \cite{2016MNRAS.455..343P}, \cite{2017JGeod..91..767X}, \cite{2018JGRB..12310162A}. Single-band measurements are sensitive to the brightest part position itself; thus, they are affected by the full magnitude of the core position bias addressed in this paper.

\begin{figure}
  \centering
  \includegraphics[width=0.96\columnwidth]{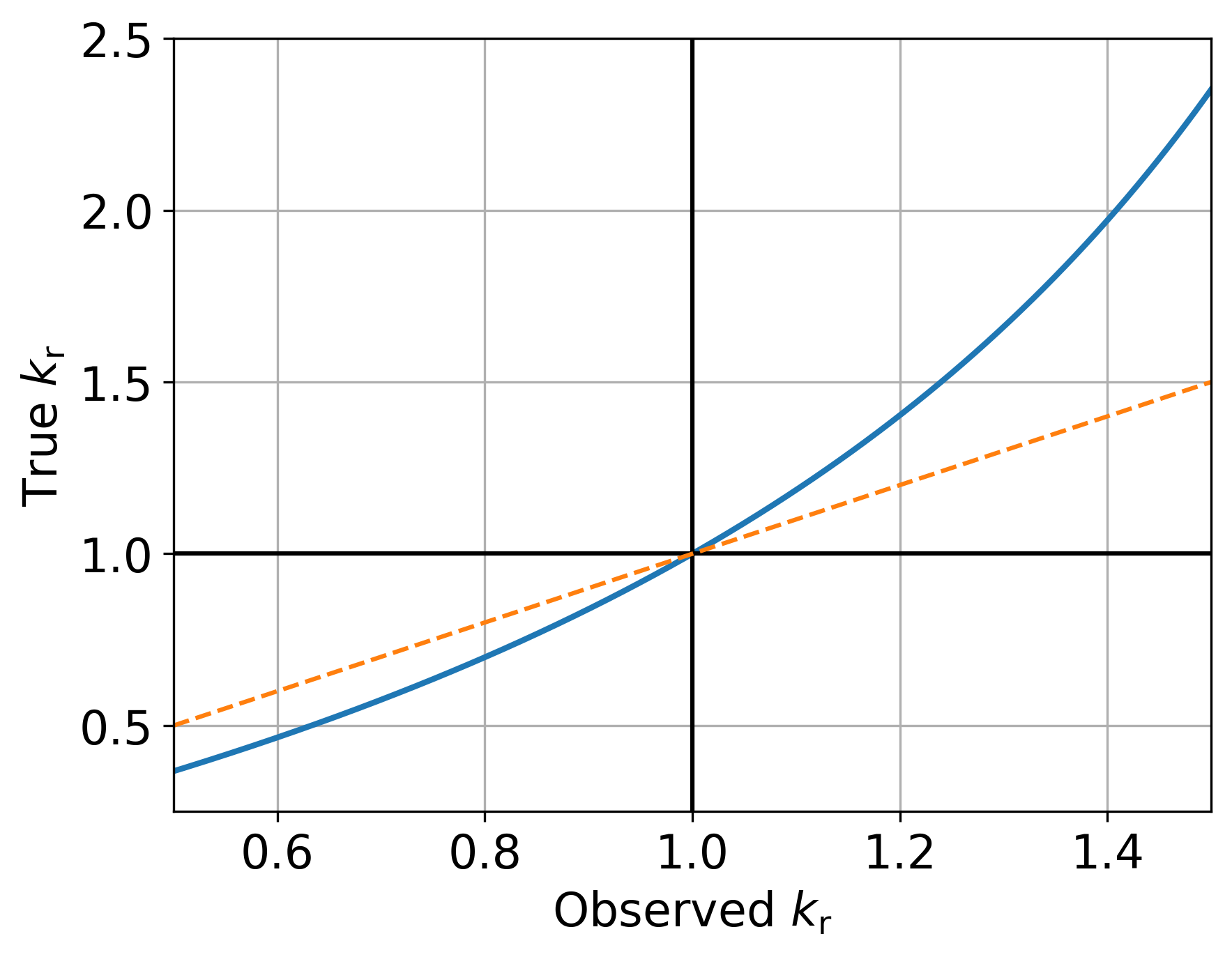}
      \caption{Blue line shows the dependence of the observed vs the true $k_{\rm r}$. The line of equality is shown by the orange color.}
         \label{fig:k_r_true_vs_observed}
  \end{figure}

Commonly used multi-band astrometry is sensitive to the frequency dependence of the core offset \citep{2009A&A...505L...1P}: the more it departs from $r \sim \nu^{-1}$, the larger the position measurement errors are. Such departures often arise due to intrinsic physical properties of AGNs: see \cite{1998A&A...330...79L} for theoretical modelling and \cite{2011MNRAS.415.1631K,2012MNRAS.420..542A,2014MNRAS.437.3396K,2016A&A...590A..48K} for specific observational results. \cite{2019MNRAS.485.1822P} have shown that this frequency dependence is also affected by strong flares which commonly happen in AGN jets. Our analysis suggests that systematic errors of astrometric positions in all these cases are effectively amplified. Indeed, as illustrated in Fig.~\ref{fig:k_r_true_vs_observed}, any departure of the exponent in $r \sim \nu^{-k_{\rm r}}$ from $k_{\rm r}=1$ gets magnified as an effect of the core shift measurement bias. Quantitative estimates of this influence on astrometric positions depend on specific methods used and are out of the scope of this paper.

In principle, the effects of the extended source structure on astrometric measurements are independent from the effects of the frequency-dependent core position \citep{2009A&A...505L...1P}. However, the extended structure is often described with a simple model like a set of Gaussian components \citep{1990AJ.....99.1309C}. The determination of these components parameters can also be degraded by biases of the same nature as those studied here.



\section{Summary}
\label{sec:conclusions}

In this paper, we examined and found significant biases in measurements of the frequency dependent core shift effect in AGN jets and suggested a method to compensate for them.
To estimate the model approximation bias, we performed simulations using the artificially generated multi-frequency VLBI data sets based on the brightness distributions derived from the \cite{1979ApJ...232...34B} jet model and the parameters of the complete MOJAVE sample. We employed two approaches to modelling the core with a Gaussian template: fitting source structure with a number of Gaussians and fitting the core with a single Gaussian after subtracting the extended emission.
Our results show that the estimated core positions and core shifts are positively biased.
For example, for the 15.4-8.1~GHz frequency interval, the typical bias is $\approx$ 0.02--0.08 mas, i.e. comparable to the core shift random error. In our simulated sample, the overestimation factor reaches the highest value of two for the sources with large apparent jet opening angles. The observed-to-true core shift ratio depends mostly on the true core shift in units of the beam size, which can be employed to compensate for the bias.
The exponent $k_{\rm r}$ of the core shift frequency dependence $r(\nu) \sim \nu^{-k_{\rm r}}$ is biased for the $k_{\rm r} \ne 1$ case. In fact, any departure from the canonical value $k_{\rm r} = 1$ is underestimated.
We also show that mixing two different definitions of the VLBI core~--- the position of the $\tau=1$ surface and the position of the maximal intensity~--- could result in biased estimates of the magnetic field.
Our results imply that the estimates of the jet magnetic fields made using measured values of the core shift could be positively biased up to an order of a magnitude depending on the estimation method.
The apparent speed estimated from the core shift and the variability time lags as well as the distances of the core from jet origin obtained from the core shift might also be overestimated to some degree.
Estimates of the jet geometry obtained from the core shift and size measurements are biased.
They become consistent with a parabolic jet shape after correcting for the bias.

\section*{Data Availability}

The datasets were derived from sources in the public domain: \url{http://www.physics.purdue.edu/astro/MOJAVE/allsources.html}, \url{http://astrogeo.org/vlbi\_images/}.

\section*{Acknowledgements}

Authors thank the anonymous referee for comments and suggestions which helped to significantly clarify the paper.
Authors thank Eduardo Ros for extensive comments on the manuscript and Elena Bazanova for language corrections.
The Russian Science Foundation grant 16-12-10481 has supported the core shift systematic errors analysis.
The Russian Basic Research Foundation grant 19-32-90141 has supported estimates of the AGN inner jet parameters.
This research has made use of data from the MOJAVE database that is maintained by the MOJAVE team \citep{2018ApJS..234...12L}.
This research has made use of NASA's Astrophysics Data System.
This research has made use the Astrogeo VLBI FITS image database.

This research made use of \textit{Astropy}, a community-developed core Python
package for Astronomy \citep{2013A&A...558A..33A}, \textit{Numpy}
\citep{numpy}, \textit{Scipy} \citep{scipy}, \textit{scikit-learn} \citep{scikit-learn} and \textit{PyMC3} \citep{10.7717/peerj-cs.55}. \textit{Matplotlib} Python package \citep{Hunter:2007} was used for generating all plots in this paper.




\bibliographystyle{mnras}
\bibliography{litr} 

\bsp	
\label{lastpage}
\end{document}